\documentclass{tlp}

\usepackage{amsmath}
\usepackage{amssymb}
\usepackage{enumerate}
\usepackage{tikz}
\usepackage{stmaryrd} 
\usepackage{ifthen} 
\usepackage{mathtools} 
\usepackage{cmll} 
\usepackage{aliascnt}
\usepackage{cleveref} 

\newcommand\ie{i.e.\@}
\newcommand\abbr[3]{\ifthenelse{\equal{#1}{}}{#2}{#1}#3}

\newcommand\wrt{{w.r.t.}}
\newcommand\st{{s.t.}}

\newcommand{\lra}{\longrightarrow}

\newcommand{\ra}{\rightarrow}

\newcommand\rlh\rightleftharpoons

\newcommand{\vdashv}{{\dashv\vdash}}
\newcommand{\vdashc}{\vdash_{\C}} 
 
\newcommand{\Vdashc}{\Vdash_{\C}}

\newcommand{\vdashvc}{{\vdashv_{\C}}}

\newcommand{\C}{{\cal C}}
\newcommand{\E}{{\cal E}}

\newcommand{\D}{{\cal D}}

\newcommand\R{{\cal R}}

\newcommand\oac[2]{{\mathcal O}^{\text{$#1$}}({#2})}

\newcommand\bfvec[1]{{\bf #1}}
\newcommand\Prog{{\mathcal P}}
\renewcommand\Prog{{\mathcal P}}

\newcommand\Newtheorem[2]{
  \newaliascnt{#1}{theorem} 
  \newtheorem{#1}[#1]{#2}  
  \aliascntresetthe{#1} 
}

\crefname{figure}{Fig\@.}{Fig\@.}
\crefname{theorem}{Thm\@.}{Thm\@.}
\crefname{proposition}{Prop\@.}{Prop\@.}
\crefname{section}{Sect\@.}{Sect\@.}
\crefname{subsection}{Sect\@.}{Sect\@.}
\crefname{example}{Ex\@.}{Ex\@.}
\crefname{corollary}{Coro\@.}{Coro\@.}

\newcommand\mathmodcond[2][]{\({#2}\)}

\newtheorem{theorem}{Theorem}[section]
\Newtheorem{lemma}{Lemma}
\Newtheorem{proposition}{Proposition}
\Newtheorem{corollary}{Corollary}
\Newtheorem{definition}{Definition}
\Newtheorem{example}{Example}
\Newtheorem{remark}{Remark}

\renewcommand\cref[1]{\Cref{#1}}

\newcommand\fv{{\rm fv}}
\newcommand\Vars{{\mathcal V}}
\newcommand\Sigmaf{\Sigma_f}
\newcommand\Sigmac{\Sigma_c}

\newcommand\askBasis[2]{{#1 \rightarrow #2}}
\newcommand{\ask}[3][nada]{
  \ifthenelse{\equal{#1}{nada}}
    {\askBasis{#2}{#3}}
    {{\forall #1 (\askBasis{#2}{#3})}}}
\newcommand{\Ask}[3][nada]{
  \ifthenelse{\equal{#1}{nada}}
    {\askBasis{#2}{#3}}
    {{\forall #1 \left(\askBasis{#2}{#3}\right)}}}

\newcommand\bangaBasis[2]{{#1 \Rightarrow #2}}
\newcommand{\banga}[3][nada]{
  \ifthenelse{\equal{#1}{nada}}
    {\askBasis{#2}{#3}}
    {{\forall #1 (\bangaBasis{#2}{#3})}}}

\newcommand{\hide}[2][]{
   \ifthenelse{\equal{#1}{}}
   {{#2}}
   {{\exists} {#1} {#2}}}

\newcommand{\para}{|}

\newcommand{\tell}[1]{\overline{#1}}

\newcommand\omBasis[1]{{\overline {#1}}}
\newcommand\om[2][]{
  \ifthenelse{\equal{#1}{}}
    {\omBasis{#2}}
    {{({#1})\omBasis{#2}}}}

\newcommand\im[2][]{
  \ifthenelse{\equal{#1}{}}
    {#2}
    {{({#1}){#2}}}}

\newcommand\som[1]{TO CHANGE}

\newcommand\lram[1]{\xrightarrow{\footnotesize #1}}

\newcommand\Lram[1]{\xRightarrow{\footnotesize #1}}

\newcommand\lramc[1]{\lram{#1}_{\mathcal C}}
\newcommand\lraomc[2][]{\lramc{\om[#1]{#2}}}
\newcommand\lraimc[2][]{\lramc{\im[#1]{#2}}}
\newcommand\Lramc[1]{\Lram{#1}_{\mathcal C}}

\newcommand\bv{{\rm ev}}

\newcommand\linfer[3][]{ {\small $\displaystyle {\frac{#2}{#3}}$%
\ifthenelse{\equal{#1}{}}{}{{ \it \small (#1)}}%
} }
\newcommand\laxiom[2][]{ {\small $\displaystyle {#2}$%
\ifthenelse{\equal{#1}{}}{}{\, {\it \small (#1)}}%
} }

\newcommand\equivc{\equiv_\C}

\newcommand\fork{\texttt{\small\tt frk}}
\newcommand\eat{\texttt{\small\tt eat}}

\newcommand\lrapm[1]{\xrightarrow{\footnotesize #1}_\pi}

\newcommand\lramcpi[1]{\lram{#1}_{\mathcal C_\pi}}

\newcommand\trans[3][]{{{\llbracket {#3} \rrbracket}_{#2}^{#1}}}
\newcommand\transpi[1]{\trans{\pi}{#1}}
\newcommand\Cpi{{\C_\pi}}

\newcommand\Cc{{\mathcal{C}_c}}

\newcommand\Vdashccc{\vdash_{\C}}

\newcommand\false{\textbf{false}}
\newcommand\true{\textbf{true}}

\newcommand\BB{{\mathbb B}}
\newcommand\CC{{\mathbb C}}
\newcommand\DD{{\mathbb D}}
\newcommand\EE{{\mathbb E}}

\newcommand\GG{{\mathbb G}}
\newcommand\HH{{\mathbb H}}
\newcommand\KK{{\mathbb K}}

\newcommand\simplif{~{\Longleftrightarrow}~}
\newcommand\propag{~{\Longrightarrow}~}

\newcommand\default[2]{\ifthenelse{\equal{#1}{}}{#2}{#1}}
\newcommand\state[3]{\langle \default{#1}{\emptyset};\default{#2}{\true}
;\default{#3}{\emptyset}
 \rangle}

\newcommand\CT{{\mathcal{CT}}}
\newcommand\at{{\;@\;}}

\newcommand\prth[1]{{\left( {#1} \right)}}

\newcommand\eq{\texttt{eq}}

\newcommand\fun{\text{t}}
\newcommand\ratr{\rightarrowtriangle}

\title[Observational equivalences for linear logic CC languages] {Observational
  equivalences for linear logic concurrent constraint languages%
\thanks{A version of the paper including the proofs is available as
  technical report (Haemmerl\'e 2011).}
}
\author[R\'emy Haemmerl\'e]{R\'emy Haemmerl\'e\\
Technical University of Madrid}

\begin{document}
\maketitle

\nocite{Haemmerle11clipB}

\begin{abstract}
   Linear logic Concurrent Constraint programming (LCC) is an extension
  of concurrent constraint programming (CC) where the constraint
  system is based on Girard's linear logic instead of the classical
  logic.  In this paper we address the problem of program equivalence
  for this programming framework. For this purpose, we present
  a structural operational semantics for LCC based on
  a label transition system and investigate different notions of
  observational equivalences inspired by the state of art of process
  algebras.  Then, we demonstrate that the asynchronous
  $\pi$-calculus can be viewed as simple syntactical restrictions of
  LCC. Finally we show LCC observational equivalences can be
  transposed straightforwardly to classical Concurrent Constraint
  languages and Constraint Handling Rules, and investigate the
  resulting equivalences.
\end{abstract}
\section{Introduction}
\label{section:introduction}

The class of Concurrent Constraint languages (briefly,
CC)~\cite{SR90popl} was introduced as a generalization of concurrent
logic programming~\cite{Maher87iclp} with constraint logic
programming~\cite{JL87popl}. Nonetheless it has strong similarities
with more classical models of concurrency such as the Calculus of
Communicating Systems (CCS), the Chemical Abstract Machine (CHAM), or
the $\pi$-calculus. %
For example, its semantics has been originally expressed by process
algebras similar to CSS~\cite{SR90popl} or later in the style of the
CHAM~\cite{FRS01ic}. Furthermore, it generalizes Actor
model~\cite{KS90oopsla} and possesses the phenomenon of channel
mobility of the $\pi$-calculus~\cite{LM92mfcs}.

Nonetheless, any CC language differs from the usual models of
concurrency because it relies on a constraint system for specifying
relationship (entailment) between messages (constraints), which
confers to it a ``monotonic'' essence. %
 Indeed, in CC, processes can only add
information by posting constraints or checking that enough information
is available to entail a {\em guard}. %
{\em Linear logic CC} languages (briefly, LCC)~\cite{SL92xerox} have
been introduced as a generalization of CC in which processes can
consume information by means of the ask operation, hence breaking the
monotonicity of CC. The main idea of this extension is to view the
constraint system as Girard's linear logic~\cite{Girard87tcs} theory
instead of classical logic theory. It results in a simple framework
that unifies constraint programming and asynchronous process algebras.

Since the beginning of the nineties, the semantics
foundation of LCC has been well studied (See
for instance~\cite{BBP97coord,RF97csl,FRS01ic,HFS07fsttcs}), but
surprisingly the formal comparison with classical models of
concurrency has received little attention. Indeed, during the same
period, the use of constraints in the context of concurrency seems to
have received more than a little
attention. For instance, the fusion calculus \cite{PV98lics}
introduced at the end of the nineties can be viewed as a
generalization of the $\pi$-calculus with unification
constraints. Several hybrid process algebras with constraint
mechanisms have also been proposed 
(See for example  \cite{DRV98clei,GP00cl,BM07esop}).%

In this paper, we investigate observational equivalence for LCC. %
Here, we understand observational equivalence in a broad sense: two
processes are observationally equivalent if, in any environment, an
external observer cannot possibly tell the difference when one process
is unplugged and the other one plugged in. %
In order to provide a relevant instantiation for this intuitive
definition, it is necessary to take into account the execution
paradigm in which the processes will be considered. %
Indeed, in CC frameworks there typically exist two possible execution
paradigms: the ``backtracking'' paradigm (from logic programs), which
allows reversible executions, and the ``committed choice'' paradigm
(from process algebras), which does not.%
In the following, we propose the {\em may testing equivalence} and the
{\em barbed congruence}, as natural instances of observable
equivalence for LCC when considered in these respective paradigms. We
propose also the {\em logical equivalence} and the {\em labelled
  bisimulation} that will provides simpler characterization for the
two former notions.%

In order to define such equivalences, we will look at LCC from a
point of view slightly different from the classical one: Here
constraints are not posted into a central blackboard anymore, but they
are processes that can migrate, merge, and emit as message
a part of the information they represent; meanwhile, ask processes
just wait for messages that ``logically'' match
their guards. Hence, it is possible to express the operational
semantics of LCC by an elegant labeled transition system (briefly,
LTS).
We then show that the asynchronous $\pi$-calculus can be viewed as
a sub-calculus of LCC, and
  that the usual $\pi$-calculus observational
equivalences are particular
instances of the ones of LCC.
Finally, we investigate particular properties of LCC observational
equivalences, when they are transposed into
classical CC and Constraints Handling Rules (CHR).

\section{A process calculi semantics for Linear Logic CC}
\label{section:lcc}

In this paper, we assume given a denumerable set $\Vars$ of variables,
a denumerable set $\Sigmac$ of predicate symbols (denoted by
$\gamma$), and a denumerable set $\Sigmaf$ of function and constant
symbols.  First order terms built from $\Vars$ and
$\Sigmaf$ will be denoted by $t$. Sequences of variables or terms will
be denoted by bold face letters such as $\bfvec x$ or $\bfvec t$. For an
arbitrary formula $A$, $\fv(A)$ denotes the set of free variables
occurring in $A$, and $A[\bfvec x\backslash\bfvec t]$ represents $A$ in
which the occurrences of variables $\bfvec x$ have been replaced by
terms $\bfvec t$ (with the usual renaming of bound variables, avoiding
variable clashes).

\subsection{Syntax}
\label{subsection:lcc:sytax}

In this section, we give a presentation of LCC languages where
declarations are replaced by replication of guarded processes.
Indeed, 
replicated asks generalize usual declarations to closures with
environment represented by the free variables in the ask
\cite{HFS07fsttcs}.  In LCC, we distinguish four syntactical categories
as specified by the following grammar:
\begin{center}
\newcommand\LOCALconstraints{c ::= {\bf 1} \mid {\bf 0} \mid {\gamma}(\bfvec t) \mid c \otimes c \mid \exists x.c \mid \oc c
& \qquad \qquad & (\text{\it constraints})}
\newcommand\LOCALactions{\alpha ::= \tau  \mid \im{c} \mid \om[\bfvec x]{c}
&\qquad \qquad  & (\text{\it LCC-actions})}
\newcommand\LOCALguards{G ::= \ask[\bfvec x]{c}{P} \mid G + G  &\qquad \qquad  & (\text{\it LCC-guards})}
\newcommand\LOCALprocesses{P ::=  
\tell{c} \mid P \para P \mid \hide[x]{P}  \mid \oc G \mid G & & (\text{\it LCC-processes})}
\small \hfill $\begin{array}{lcr}
\LOCALconstraints\\ \LOCALactions \\ \LOCALguards \\ \LOCALprocesses
\end{array}$ \hfill \null
\end{center}

Constraints are formulas built from terms, constraint symbols, and the
logical operators: ${\bf 1}$ (true), ${\bf 0}$ (false), the conjunction
$(\otimes)$, the existential quantifier $(\exists)$, and the modality
$(\oc)$. %
The three kinds of actions are the {\em silent action} $\tau$, the {\em input
  action} $\im{c}$, which represents a constraint for which a process
waits, and the {\em output action} $\om[\bfvec x]{c}$ ($\bfvec x$ being
the variables {\em extruded} by the action), which represents the
constraint posted by a process.
The order of the extruded variables in an output message is irrelevant,
hence if $\bfvec y$ is a permutation of the sequence $\bfvec x$, we will
consider $\om[\bfvec x]{c}$ equal to $\om[\bfvec y]{c}$.  In
LCC-processes, an overlined constraint $\om{c}$ stands for {\em
  asynchronous tell}, $\para$ for {\em parallel composition},
$\exists$ for {\em variable hiding}, $\ra$ for {\em blocking ask}, $+$
for guarded choice, and $\oc$ for {\em replication}. %
As one can see, the syntax for LCC-processes does not include specific
construction for the null process. %
Indeed, this latter can be emulated by the trivial constraint
$\tell{{\bf 1}}$, which represents no information.

For convenience, if $\bfvec x$ is empty, we will abbreviate $\ask[\bfvec
x]{c}{P}$ and $\om[\bfvec x]{c}$ as $\ask{c}{P}$ and $\om{c}$, respectively. %
$\hide[\bfvec x]{A}$ will be a notation for $\hide[x_1]{\dots
  \hide[x_n]{A}}$ if $A$ is a constraint or an LCC-process and $\bfvec
x$ is the sequence of variables $x_1\dots x_n$. %
Moreover, for any finite multiset of processes
  $\{P_1, \dots, P_n\}$,
we will
use
  $\Pi_{i=1}^n P_i$
as abbreviations
for
  $P_1 \para \cdots \para P_n$.
As usual, the existential and universal quantifiers in constraints and
LCC-processes are considered as variable binders. Conventionally, we
consider the variables $\bfvec x$ as free in any action of the form
$\om[\bfvec x]{c}$.  We use $\bv(\alpha)$ as an abbreviation for the
extruded variables of $\alpha$ (\ie\ $\bv(\alpha) = \bfvec x$, if
$\alpha$ is an action
of the form $\om[\bfvec x]{c}$, $\bv(\alpha) =
\emptyset$ otherwise).

LCC languages are parametrized by {\em a (linear) constraint system},
which is a pair $(\C,\Vdashc)$ where $\C$ is the set of all
constraints and $\Vdashc$ is a subset of $\C\!\times\!\C$ which defines
the non-logical axioms of the 
system. %
For a given constraint system $(\C,\Vdashc)$, the entailment relation
$\vdashc$ is the smallest relation containing
$\Vdashc$ and closed by the rules of intuitionistic linear
logic%
. We will use the notation $A \vdashvc B$ to mean
that both sequents $A \vdashc B$ and $B \vdashc A$ hold. %

In this paper, we are interested in studying classes of LCC processes
obtained by syntactical restrictions on the constraints that
they can use. These restrictions will
  simulate the power of the observer in LCC sub-calculi and/or the
  visibility limitations imposed by ad-hoc scope mechanisms such as
  module systems. In practice, they will be
  specified by
means of two subsets of $\C$, that will
limit the possible constraints a process can respectively ask or tell.
Formally for all subsets $\D$ and $\E$, we say
that a process $P$ is {\em $\D$-ask restricted} (resp. {\em $\E$-tell
  restricted}) if it is obtained by the grammar for processes where
any ask $\ask[\bfvec x]{c}{P}$ (resp. any tell $\overline c$)
satisfies $(\exists \bfvec x.c)\in \D$
(resp. $c \in \E$). More generally, we say that $P$
is a $\D\E$-process if $P$ is both $\D$-ask and $\E$-tell restricted.

\subsection{Operational semantics}
\label{subsection:lcc:operational}

{
\newcommand\LOCALcinrule{
\laxiom[$\C$-in]{ \tell{{\bf 1}} \lraimc{c} \tell{c}}}
\newcommand\LOCALcoutrule{\linfer[$\C$-out]{
\begin{array}{c}
c \vdashc \exists \bfvec x(d \otimes e) \quad
\exists \bfvec x d \vdashc \exists {\bfvec x'} d'  %
\quad \bfvec {xx'} \cap \fv(c) = \emptyset
 \\
 c' \vdashc  \exists \bfvec x (d'    \otimes e)  \text{ is a most general choice }
\end{array} 
}{\tell{c} \lraomc[\bfvec x']{d'} \tell{e}}}
\newcommand\LOCALccomprule{\linfer[$\C$-comp]
{P\lramc{\alpha} P' \quad \bv(\alpha) \cap \fv(Q)\!=\!\emptyset}
{P\para Q \lramc{ \alpha} P'\para Q}}
\newcommand\LOCALcsyncrule{\linfer[$\C$-sync]
{\begin{array}{c}
c \vdash_\C \exists \bfvec y( d[\bfvec x\backslash \bfvec t] 
     \otimes e ) \quad 
\bfvec y \cap \fv(c, d, A) = \emptyset
 \\ \exists \bfvec y (d[\bfvec x\backslash \bfvec t] 
     \otimes e)  \text{ is a most general choice}\end{array} 
}
{\tell{c} \para \ask[\bfvec x]{d}{A} \lramc{\tau} \exists \bfvec y.(A[\bfvec x\backslash \bfvec t] \para \tell{e})}}
\newcommand\LOCALcoutexrule{\linfer[$\C$-ext]
  {P \lraomc[\bfvec x]{c} Q}
  {\hide[y]{P} \lraomc[y \bfvec x]{c} {Q}}}
\newcommand\LOCALcrestrule{\linfer[$\C$-rest] 
  {P \lramc{\alpha} Q  \quad  y \notin\! \fv(\alpha) }
{\hide[y]{P} \lramc{\alpha} \hide[y]{Q}}}
\newcommand\LOCALcongrule{ \linfer[cong]{P\! \equiv \! P' \;\; P' \! \lram{\alpha}\! Q' \;\; Q'\! \equiv\! Q}
{P  \lram{\alpha} Q  }}
\newcommand\LOCALsumrule{\linfer[sum]{P \para G  \lram{\alpha} Q }{P \para (G \! +\! G')  \lram{\alpha} Q }}
\begin{table}
\vspace{-1mm}
\tiny\center
  \begin{tabular}{c}\hline\\\\
\LOCALcongrule \hfill 
\LOCALsumrule
 \\\\ 
\LOCALccomprule $\qquad$ $\qquad$ 
\LOCALcrestrule
 \\\\
\LOCALcoutrule \hfill
\LOCALcoutexrule 
 \\\\
\LOCALcsyncrule \hfill 
\LOCALcinrule
\\\\ 
 \hline
  \end{tabular}
\caption{Labeled transition system for Linear Logic CC}
\label{tab:lcc:lts}
\end{table}
}

In \cref{tab:lcc:lts}, we define, for a given constraint system
$(\C, \Vdashc)$, the operational semantics of LCC by
means of an LTS.  As usual, in process
algebras this semantics uses a structural congruence.  This
congruence, noted $\equivc$, is defined as the smallest equivalence
satisfying $\alpha$-renaming of bound variables, commutativity and
associativity for parallel composition, summation, and the following
identities:
\begin{center}
\tiny
\begin{tabular}{lclclclcl}
  & \small ${P \para \tell{{\bf 1}} \equiv_\C P}$ && %
  \small ${\hide[z]{\tell{{\bf 1}}} \equiv_\C \tell{{\bf 1}}}$  &&%
  \small ${\hide[x]{\hide[y]{P}} \equiv_\C \hide[y]{\hide[x]{P}}}$&&
  \small ${\oc P \equiv_\C P \para \oc P }$%
  & \\
  \\  & 
  \small$\displaystyle\frac{c \otimes d \vdashvc e}{\tell{c} \para \tell{d} \equiv_\C
    \tell{e}}$  &&%
  \small$\displaystyle\frac{P \equiv_\C P'}{P \para Q \equiv_\C P' \para Q}$ &&%
  \small$\displaystyle\frac{z \notin \fv(P)}{ P \para \hide[z]{Q} \equiv_\C
    \hide[z]{(P \para Q)}}$  &&%
  \small$\displaystyle\frac{P\equiv_\C P'}{\exists x.P \equiv_\C \exists x.P'}$ &%
\end{tabular}
\end{center}

The side condition ``$c \vdashc \exists \bfvec y(d[\bfvec x\backslash
\bfvec t] \otimes e)$ is a most general choice'' is a reasonable
restriction,
that guarantees the transition does not weaken constraints within a
process as can do the logical entailment (For instance we want to
avoid entailment such as $!c \vdashc c \otimes {\bf 1}$).  It can be
defined as: For any constraint $e'$, all terms $\bfvec t'$ and all
variables $\bfvec y'$ if $c \vdashc \exists \bfvec y'( d[\bfvec x\backslash
\bfvec t'] \otimes e')$ and $\exists \bfvec {y'}e'\vdashc \exists \bfvec
{y}e$ hold, then so do $\exists \bfvec {y}d[\bfvec x\backslash \bfvec t]
\vdashc \exists \bfvec {y'}d[\bfvec x \backslash \bfvec t']$ and $\exists
{\bfvec y}e \vdashc \exists \bfvec {y'}e'$.
In the constraint systems we will consider in this article, such a
deduction is always possible.

 The notion of {\em weak transition} is defined classically:
\begin{center}\small
$
(P\Lramc{\tau} Q) \; \stackrel{def}{\Longleftrightarrow} \;%
( P \lramc{\tau}^* Q) \qquad \qquad
( P\Lramc{\alpha} Q) \; \stackrel{def}{\Longleftrightarrow} \;
(P \lramc{\tau}^* \lramc{\alpha} \lramc{\tau}^* Q) ~~ \text{ (for $\alpha
  \not = \tau$)}
$  
\end{center}

In the asynchronous context of this paper, it seems
natural to restrict the observation to outputs. As argued by 
\citeN{ACS98tcs}, the intuition is that an observer cannot know that a
message he has sent has been actually received.
Moreover, since an observer has no way of knowing if the execution of a
particular process is terminated unless he receives a programmed
acknowledgment, we will disregard classical (L)CC observables which deal
with termination such as success stores \cite{SRP91popl,FRS01ic}, and 
consider only accessible constraints \cite{HFS07fsttcs}.
Formally for any set $\D \subset \C$, the set of {\em $\D$-accessible
  constraints} for a process $P$ is defined as:
\begin{center}
\small\hfill
$
\oac{\D}{P} = \left\{ (\exists \bfvec x.c) \in \D \mid 
\text{ there exists $P'$ such that }P\Lramc{\tau} \exists \bfvec x.(P' \para c) 
 \right\}
$ \hfill\null
\end{center}

The semantics we propose has important links with the one defined by
\citeN{BBP97coord} but it is in some important aspects more general. In
particular, the language we consider provides replication and explicit
operators for both universal and existential quantifications, all of
which are important features.  Indeed, on the one hand replication and
existential quantification are crucial to internalize declarations and
closures in processes \cite{HFS07fsttcs}; while, on the other hand
universal quantification cannot be emulated by tell processes in every
constraint system, especially linear ones \cite{FRS01ic}.
Another difference is that our system uses the {\em asynchronous
    input} rule as initially proposed by \citeN{HY95tcs} for the
  $\pi$-calculus.  This rule, which allows an observer to do
  any input action at any time, is not designed to be observed
  directly but rather to simplify bisimulation-based  definitions 
  within asynchronous frameworks~\cite{ACS98tcs}.

\begin{example}[Dining philosophers]\label{example:lcc:philos} 
  As suggested by \citeN{BBP97coord}, 
  the dining philosophers problem has an extremely simple solution in
  LCC.  Here is an adaptation of the
  solution proposed by \citeN{RF97csl}.  The atomic constraints are
  $\fork(i)$ and $\eat(j)$ for $i,j\in \mathbb N$%
  , and $\vdashc$ is the trivial entailment relation. %
  Assuming the following encoding for the i$^\text{th}$ philosopher
  among $n$, a solution for the problem
  consists of the process {\small $\Pi_{i=0}^{n-1} \left(P^n_i \para
      \tell{\fork(i)}\right)$}.
\begin{center}
\small
$
P_i^n =  \oc \left(\fork(i) \otimes \fork(i \text{+} 1 \text{ mod }  n) \ra 
\left(\tell{\eat(i)}  \para \eat(i) \!\ra\! \left(\tell{\fork(i)} \para \tell{\fork(i \text{+} 1})\right)\right)\right)
$
\end{center}
This solution suffers neither deadlock nor starvation problems: the
system can always advance to a different state, and at least one
philosopher will eventually eat.
\end{example}

\subsection{Logical semantics}
\label{subsection:lcc:logical}
In this section, we show that the results of logical semantics from
LCC~\cite{FRS01ic,HFS07fsttcs} can be shifted to the version of LCC we
propose in this paper. %
It will provide us with a powerful tool to reason about processes. %
It is worth noting that the logical semantics proposed here is
slightly different from the usual one, since it uses an additional
conjunction with $\top$.  As shown by the next theorem,
this modification is harmless when regarding accessible constraints,
but yields a more relevant
notion of
equivalence. (Refer to the discussion in
\cref{subsection:equivalences:logical}.) Note the conjunction with
$\top$ is not necessary in case of translation of a parallel
composition and hiding, since it commutes with $\otimes$ and
$\exists$ (\ie\ {\small
  $(A \otimes \top) \otimes (B \otimes \top) \vdashvc A \otimes B
  \otimes \top$} and {\small $\exists x(A \otimes \top) \vdashvc
  \exists x(A) \otimes \top$}).

\begin{definition}
  \label{def:lcc:logique:traduction}
 Processes are translated into 
linear logic formulas as follows:
\begin{center}\small
$\begin{array}{rclrclrcl} 
\tell{c}^\dagger   &=&  c  \otimes \top & %
(P \para Q)^{\dagger}  &=&  P^\dagger \otimes Q^\dagger &  
  (P + Q)^\dagger  &=& (P^\dagger \with Q^\dagger) \otimes \top \\
(\oc P)^\dagger   &=& \oc (P^\dagger) \otimes \top & %
(\hide[x]P)^\dagger   &=& \exists x P^\dagger  & %
(\ask[\bfvec x]{c}{ P})^\dagger &=&\forall \bfvec x  ( c   \multimap  P^\dagger) \otimes \top
\end{array}$
\end{center}
\end{definition}

\begin{theorem}[Logical semantics]
 \label{LccLogical}
For any process $P$ and any set $\D$ of linear constraints, 
\mathmodcond{\oac{\D}{P} = \left\{ d \in \D \mid P^\dagger \vdashc d ^\dagger \right\}.}
\end{theorem}

\section{Observational equivalence relations for Linear logic CC}
\label{section:equivalences}

In this section, we propose some equivalence relations for
LCC-processes.

An important property of processes related by equivalences is their
dependence on the environment. More precisely, two equivalent
processes must be indistinguishable by an observer in any context
(i.e\@. equivalences must be congruences). Formal {\em contexts},
written $C[\,]$, are processes with a special constant $[\,]$,
the hole.  Putting a term $P$ into the holes of a
context $C[\,]$ gives the term noted $C[P]$. In practice, we define all
our congruences for {\em evaluation contexts~\cite{FG05jlap}}, a
particular class of contexts 
  where the hole occurs
  exactly once and not under a guard nor a replication. These
  contexts, also called static contexts~\cite{Milner89prentice},
describe environments that can communicate with an observed process
and filter its messages but can neither substitute variables of the
process nor replicate it.
In this paper, without explicit statement of the contrary, all
congruence properties will refer to these contexts only. In
particular, we will use the terminology ``{\em full congruence}'' to
refer to the congruence with respect to arbitrary contexts. 
  In the framework of LCC, {\em $\D\E$-contexts} and {\em
    $\D\E$-congruence} will refer to evaluation contexts and
  congruence built from $\D\E$-processes.

\subsection{Logical equivalence}
\label{subsection:equivalences:logical}

Strictly speaking, the first notion of equivalence we
consider is not observational, but stems naturally from the logical
semantics of the language. Indeed, the logical semantics ensures that
processes with logically equivalent translations
have the same accessible constraints. This notion of
  equivalence is specially interesting since it can be proved using
  automated theorem provers such as llprover~\cite{Tamura98Kobe}. 

\begin{definition}[Logical equivalence]
  The {\em
    (weak)} {\em logical equivalence} on
  LCC-processes is defined as:
  \begin{center}\small$
   P \multimapboth_\C Q \; \stackrel{def}{\Longleftrightarrow} 
  \;P^\dagger  \vdashvc Q^\dagger$
  \end{center}
\end{definition}

We call this equivalence ``weak'' because it is strictly less
  discriminating than the
one we would obtain using usual logical
semantics of LCC.
Nonetheless, the present definition is more relevant since it does not
distinguish Girard's exponential connective, noted $\oc$ in Linear
Logic, from Milner's replication, noted also $\oc$ in process
algebras. Indeed, for any linear logic formula $A$, $\oc A \otimes A
\otimes \top \vdashv \oc A \otimes \top$ holds, whereas $\oc A \otimes
A \vdashv \oc A$ does not.  The proposition we give next
  states that the use of
$\top$ does not break the congruence property of logical equivalence.

\begin{proposition}
\label{proposition:logical:congruence}
Weak logical equivalence is a full congruence.
\end{proposition}

\subsection{May-testing equivalence}\label{subsection:equivalences:testing}

The following equivalence relates to testing
semantics~\cite{DNH84tcs}.  We argue that this relation
  provides a canonical notion of observational equivalence for LCC if
  considered within the ``backtracking'' execution paradigm.
Indeed, it is defined as the
  largest congruence that respects accessible constraints.
For the sake of generality, we defined may-testing in a parametric way
according to input/output filters.

\begin{definition}[May-testing equivalence]
  Let $\D$ and $\E$ be two subsets of $\C$.  The {\em
    may $\D\E$-testing}, $\simeq_{\D\E}$, is the largest
  $\D\E$-congruence that respects $\D$-accessible constraints,
  formally:
  \begin{center}\small$
  P \simeq_{\D\E} Q \; \stackrel{\small def}{\Longleftrightarrow} \;
  \text{for any evaluation $\D\E$-context $C[\,]$, } \oac{\D}{C[P]} =
  \oac{\D}{C[Q]}.$
  \end{center}
\end{definition}

Quite naturally, logical equivalence implies  any may testing equivalence
relation. One can use logical semantics and
\cref{proposition:logical:congruence} to demonstrate it.  It is
worth noting that the inclusion is strict. For instance, the processes
$\ask{c}{\exists x.P}$ and $\exists x.(\ask{c}{P})$, where $x$ is free
in $P$ and not in $c$, are clearly equivalent with
respect to any may testing equivalence but are not
logically equivalent in linear logic.

\begin{example}\label{example:equivalences:testing}
  Contrary to the processes in \cref{example:lcc:philos}, the
  following implementation for the $i^\text{th}$ dining
    philosopher does not use atomic consumptions of constraint
    conjunctions:
  \begin{center}\small
$Q_i^n \! =\  \oc\! \left(\fork(i) \! \ra \! \left(\fork(i\text{+} 1 \text{ mod }  n) \!\ra\! 
\left(\tell{\eat(i)}  \para \eat(i) \!\ra\! \left(\tell{\fork(i)} \para \tell{\fork(i\text{+} 1 \text{ mod }  n)}\right)\!\right)\!\right)\!\right)$
  \end{center}
  Although the solutions built with such philosophers face 
  deadlock and starvation problems, the two implementations of
  philosopher cannot be distinguished by may-testing (\ie\ for all
  {\small $i,n\in \mathbb N$}, {\small $P_i^n
    \simeq_{\C\C} Q_i^n$}). %
  Note that in the ``backtracking'' execution paradigm there is no
  reason to distinguish such processes. Indeed, the possibility of
  reversing executions makes deadlocks invisible from an external
  point of view.
\end{example}

\subsection{Labeled Bisimulation}
\label{subsection:bisimulation}

In the framework of process algebra, bisimulation-based equivalence
relations are the most commonly used notion of equivalence.  %
Contrary to the may-testing equivalences and the barbed congruences
presented in the following, the labeled bisimulation proofs do
not require explicit context closure. Indeed, as shows
\cref{proposition:bisimulation:congruence}, congruence is not a
requirement but a derived property. %
Hence, the proofs can be established by coinduction, by considering
only few steps. %
As we have done for may-testing, our definition of bisimulation is
parametrized by input/output filters.

\begin{definition}[Labeled bisimulation]
\label{definition:bisimulation}
Let $\D$ and $\E$ be two
  subsets of $\C$. A action is {\em $\D\E$-relevant} for a process $Q$
  if it is either a silent action, or an input action in $\E$, or an
  output action of the form $\om[\bfvec x]{c}$ with $(\exists \bfvec x.c)
  \in \D$ and $\bfvec x \cap \fv(Q) = \emptyset$.
  A symmetrical relation $\R$ is a {\em $\D\E$-bisimulation} if for
  all $P$, $P'$, $Q$, $\alpha$ such that $P\R Q$, $P \lramc{\alpha}
  P'$, and $\alpha$ is $\D\E$-relevant for $Q$, there exists
  $Q$' such that $Q \Lramc{\alpha} Q'$ and $P' \R Q'$. 
  The largest $\D\E$-bisimulation is called {\em $\D\E$-bisimilarity}
  and is denoted with $\approx_{\D\E}$.
\end{definition}

\begin{theorem}
\label{proposition:bisimulation:congruence}
For all sets of constraints $\D$ and
$\E$, the $\D\E$-bisimilarity is a $\D\E$-congruence.
\end{theorem}

\subsection{Barbed congruence}
\label{subsection:equivalences:barbed}

Barbed bisimulation has been introduced by \citeN{MS92icalp} as an
uniform way to describe bisimulation-based equivalences for any
calculus. %
From the definition of observables we give in
\cref{subsection:lcc:operational}, we derive a notion of barbed
bisimulation in the standard way. %
As with many other barbed bisimulations, the obtained equivalence is
too rough. For example, no barbed bisimulation distinguishes between
processes $\tell{{\bf 1}}$ and $c\ra P$ (with $\nvdash_\C c$), which
exhibit clearly different behaviours when they are put in parallel
with a constraint stronger than $c$. For this reason, we refine our
bisimulation by enforcing congruence property following  \citeN{FG05jlap}. The resulting relation yields an
  instance of the intuitive notion of observational equivalence for
  LCC considered within the ``committed-choice'' paradigm.

\begin{definition}[Barbed congruence]
  Let $\D$ and $\E$ be two subsets of $\C$.  A symmetrical relation
  $\R$ is a {\em $\D\E$-barbed bisimulation} if for all $P$, $P'$, $Q$
  such that $P\R Q$, and $P \lramc{\tau} P'$, then there exists $Q$'
  such that $\oac{\D}{P} \subseteq \oac{\D}{Q}$, $Q \Lramc{\tau} Q'$ and
  $P' \R Q'$. %
  The {\em barbed $\D\E$-congruence}, written $\cong_{\D\E}$, is the
  largest $\D\E$-con\-gruen\-ce that is a $\D\E$-barbed bisimulation.
\end{definition}

Clearly, barbed $\D\E$-congruence is more
precise than may $\D\E$-testing equivalence. It
is worth noting that it is in general strictly distinct from logical
equivalence. For instance, $\ask{c}{\exists x.P}$ and $\exists
x.(\ask{c}{P})$ are $\C\C$-barbed congruent but not logically
equivalent, while $\ask{c}{\ask{d}{{\bf 1}}}$ and $\ask{c \otimes d}{{\bf 1}}$
are logically equivalent but not barbed congruent.
In general, direct proofs of barbed congruence are tedious since they
require explicit context closure. Fortunately, the barbed congruence
coincides with labeled bisimulation. Barbed congruence can
therefore be established by simpler proofs based on the coinductive
principle of labeled bisimulation.

\begin{theorem}
\label{theorem:bisimulation}
For all sets of constraints
$\D$ and $\E$, $\cong_{\D\E}$ and
$\approx_{\D\E}$ coincide.
\end{theorem}

\begin{example}\label{example:equivalences:bisimulation}
  The encoding of philosophers proposed in the two previous examples
  cannot be distinguished by may-testing. Nonetheless their behavior can be
  separated by barbed
  congruence. For instance, one can disprove $P_1^3
    \cong_{\C\C} Q_1^3$. The following implementation
  refines the one of %
  \cref{example:equivalences:testing} %
  by allowing a philosopher to put back the first fork he
  takes:
  \begin{center}\small
$
R_i^n \text{=}  \oc\!\left(\fork(i)  \!\ra\!  \left(\tell{\fork(i)} \!+\!
 \fork(i\text{+}1 \text{ mod } n) \!\ra\! 
     \left(\tell{\eat(i)} \para \eat(i) \! \ra \! 
       \left(\tell{\fork(i)} \para
         \tell{\fork(i\text{+}1 \text{ mod } n)}\right)\!\!\right)\!\!\right)\!\!\right)$
  \end{center}
  Although, solutions built with this latter implementation of
  philosophers still faces starvation problems, the external behaviour
  of these philosophers cannot be distinguished
    anymore from the ones of
  \cref{example:lcc:philos}, \ie\ $P_i^n \cong_{\C\C}
    R_i^n$ for any $i,n\in \mathbb N$.
\end{example}

\section{LCC a natural generalization of asynchronous calculi}
\label{section:pi}

In this section, we show that LCC language generalizes
asynchronous $\pi$-calculus. %
The asynchronous $\pi$-calculus is a variant of the $\pi$-calculus
where the emission is non-blocking.  In practice, it is obtained by a
simple syntactical restriction prohibiting output prefixing.

\medskip

We briefly recall the syntax of the asynchronous
$\pi$-calculus. Our notations and definitions are mostly standard.
For convenience, we will use a denumerable subset of LCC variables as
channel names. In this language, three syntactical categories are
distinguished as specified by the following grammar:
\begin{center} \small 
$\begin{array}{lcr}
\alpha ::= \tau \mid  \bar xy\mid \bar x(y) \mid (y)x(y)
&\qquad \qquad& (\text{\it $\pi$-actions}) \\
G ::=  \tau.P \mid  x(y).P \mid \oc P 
& & (\text{\it $\pi$-guards})\\
P ::= {\bf 0} \mid \bar x y \mid P \para P \mid \nu{x}{P}  \mid  G   & &
(\text{\it $\pi$-processes})
\end{array}$
\end{center}
A $\pi$-calculus process (or $\pi$-process for short) is one of the
following: the {\em null process} {\bf 0}, the silent prefix $\tau.P$, the {\em
  message reception } $ x(y).P$, the {\em asynchronous
  emission} $\bar xy$, the parallel composition of processes $P\para
Q$, 
 the replication of processes $\oc P$, or the {\em scope restriction} $\nu yP$.

  In this section, we assume the notion of reduction,
 which we write $\lra_\pi$, the may testing
 equivalence, which we write $\simeq_\pi$, the
 labeled bisimulation, which we write
 $\approx_\pi$, and the barbed congruence, which we write $\cong_\pi$,
 as defined by \citeN{FG05jlap}.
We propose now a very simple interpretation of the asynchronous
$\pi$-calculus into LCC following the preliminary ideas of
Soliman~\cite{soliman03inria}.

\medskip
 
\begin{definition}[LCC Interpretation of the asynchronous $\pi$-calculus]
\label{def:pure:translation}
Let $\Cpi$ be the trivial constraint system (i.e\@. a constraint
system without non-logical axioms),
 based on the predicate alphabet
$\Sigmac = \{ \gamma \}$.
The LCC-interpretation $\transpi{\;}$ of $\pi$-actions and $\pi$-processes
as 
is defined recursively as:

\vspace{-4mm}
{\small  \begin{align*} \transpi{\tau} \! =& {\tau} & \transpi{x y}
    \! =& {\gamma(x,y)} & \transpi{\bar x(y) } \! =& {\overline
      {\gamma(x,y)}} & \transpi{(y)\bar x (y)} \! =&
    {\om[y]{\gamma(x,y)}} \\
    \transpi{{{\bf 0}}} \! = & \tell{{\bf 1}} & \transpi{\bar xz} \! = & \tell{\gamma(x,z)}
    & \transpi{\tau.P} \! = & \ask{{\bf 1}\!}{\!\transpi{P}}  %
&  
 \transpi{x(y).P} \! = & \ask[y]{\gamma(x,y)\!}{\!\transpi{P}}
\\
    \transpi{\oc P} \! = & \oc \transpi{P} \!&\! \transpi{\nu x P}
    \! = & \hide[x]{\transpi{P}} \!&\!\! \transpi{P\para Q} \! =
    & \transpi{P}\para \transpi{Q} \!
&\!
  \end{align*}}
\end{definition}

It can be noted that this mapping is completely compositional and does
not need fresh names. Furthermore, the replacement of declarations by
replicated asks leads to a translation where each construct of the
$\pi$-calculus is mapped to a unique construct of
LCC. In fact, we can consider this
interpretation enforces a syntactical restriction on LCC
  processes, by allowing synchronization only on constraints of the
  form $\exists y.\gamma(x,y)$.  Formally, assuming $\D_\pi = \{ {\bf 1} \} \cup
\{ \exists y.\gamma(x,y) \mid xy\in \Vars \wedge x \not = y\}$ and
$\E_\pi = \{ {\bf 1} \} \cup \{ \gamma(x,y) \mid xy\in \Vars \}$ , the
co-domain of $\transpi{\;}$ is precisely the set of
$\D_\pi\E_\pi$-processes. Furthermore, the following results ensure
that there is a one-to-one correspondence between transitions of the
two formalisms.

\begin{theorem}
\label{theorem:pure:subcalculus}
$  P \lrapm{\tau} Q$ if and only if $
  \transpi{P}\lramcpi{\tau} \transpi{Q}.$
\end{theorem}

The theorem and the simplicity of the interpretation emphasizes that
the $\pi$-calculus is syntactically and semantically a subcalculus of
LCC.  The only transition of LCC that is not captured by the
$\pi$-calculus semantics is the simultaneous emission of messages (\ie\ a constraint of the
  form $\gamma(x_1,y_1)\otimes \dots \otimes
\gamma(x_n,y_n)$).
 We argue that observing simultaneous
emission is not relevant in asynchronous context where the observer
has no way of knowing the order in which the messages
have been emitted. 
In fact, the LCC constraint system makes messages behave
  similarly to molecules within the CHAM (\ie\ messages can combine by
  ``cooling'' and dissociate by ``heating''~\cite{BB92tcs}).

The following theorem states that may testing equivalence, labeled
bisimilarity, and barbed congruence are instances of equivalence
relations we defined for LCC. 

\begin{theorem} \label{theorem:pure:equivalences} Let $\D_\pi =
  \{ \exists y.\gamma(x,y) \mid x \in \Vars\setminus \{y\}  \}$ and
  $\D^\star_\pi \!=\! \D_\pi \cup \{ \gamma(x,y) \mid xy\in \Vars \}$.
  For all $\pi$-processes $P$ and $Q$ we have:
    \begin{enumerate}[\it (i)]
  \item $P \simeq_\pi Q$ if and only if  $\transpi{P} \simeq_{\D_\pi\C_\pi }\transpi{Q}$.
   \item $P \approx_\pi Q$ if and only if $\transpi{P}
     \approx_{\D_\pi^\star\C_\pi} \transpi{Q}$.
   \item $P \cong_\pi Q$ if and only if $\transpi{P}
     \cong_{\D_\pi\C_\pi} \transpi{Q}$.
    \end{enumerate}
\end{theorem}

\section{Observational equivalence relations for CC framework}
\label{section:cc}

\subsection{Observational equivalence relations for  classical CC }

LCC languages are refinements of CC
languages. Indeed the monotonicity of the CC store can
simply be restored with the exponential connective \(\oc\) of linear
logic, allowing duplication of hypotheses and thus avoiding constraint
consumption during synchronization \cite{FRS01ic}. Hence, all the
observational equivalence relations we defined for LCC can be
transposed effortless to classical CC. That is particularly
interesting, since few attempts can be found in the literature to
endow CC with process
equivalence techniques.

In order to further discuss properties of the resulting relations, we
will not enter into the details of a particular encoding of CC into
LCC, but just assume that the encoding of classical constraints
respects two reasonable properties.  We will say that a linear
constraint \(c\) is {\em classical} within the linear constraint system
\(\C\) (or \(\C\)-classical for short), if it can be both logically
weakened (\ie\ \(c \vdash_\C {\bf 1}\)), and deduced without
weakening the hypotheses (\ie\ for any \(d\), if \(d
\vdash_\C c \otimes \top\), then \(d \vdash_\C c \otimes d\)).  We note
\(\C_c\) the set of \(\C\)-classical constraints.  Assuming that processes
deal with classical constraints, we are able to prove some interesting
laws.  It must be underlined that, in the full generality of LCC, none of them holds.

\begin{proposition}\label{proposition:chr}
\newcommand\LOCALruleA{\displaystyle \Ask[\bfvec x]{c\!}{\!\tell{e}} \cong_{\C\C} \Ask[\bfvec x]{c\!}{\!\tell{c \otimes e}} }
\newcommand\LOCALruleB{\displaystyle \Ask[\bfvec x]{c\!}{\!\tell{c'}} \cong_{\C\C} \tell{{\bf 1}} }
\newcommand\LOCALruleC{\displaystyle \left(\Ask{d\!}{\!\Ask[\bfvec x]{e\!}{\!P}}\right) \cong_{\C} \Ask[\bfvec x]{d \otimes e\!}{\!P}}
\newcommand\LOCALruleD{\displaystyle \left(\left(\Ask{c\!}{\!P}\right)\, \para\, \left(\Ask{c\!}{\!Q}\right)\right) \cong_{\C\C}
       \left(\Ask{c\!}{\!\left(P \, \para \, Q\right)}\right)}
\newcommand\LOCALruleE{\displaystyle \left(\left(\Ask{c'\!}{\!\tell{d}}\right)\, \para\, \left(\Ask{c\!}{\!\tell{d'}}\right)\right) \cong_{\C\C} 
       \left(\Ask{c'\!}{\!d}\right)  }
\newcommand\LOCALruleF{\displaystyle \left(\left(\Ask{c'\!}{\!\tell{d}}\right)\, \para\, \left(\Ask{c\!}{\!Q}\right)\right) \cong_{\C\C}
       \left(\Ask{c'\!}{\!d}\, \para\, \left(\Ask{c \otimes d\!}{\!Q}\right)\right) }
\newcommand\LOCALruleG{\displaystyle \left(\left(\Ask{c\!}{\!\tell{d}}\right)\, \para\, \left(\Ask{d'\!}{\!P}\right)\right) \cong_{\C\C}
       \left(\Ask{c\!}{\!\left(\tell{d} \,\para \, P\right)}\right)}
\newcommand\LOCALruleH{\displaystyle  \oc\left(\Ask{c\!}{\!P}\right) \cong_{\C\C} \left(\Ask{c\!}{\!\oc P}\right)}
\newcommand\LOCALruleI{\displaystyle \left(\Ask{c\!}{\!G} + \ask{c\!}{\!H}\right)\cong_{\C\C} \left(\Ask{c\!}{\!\left(G \!+\! H\right)}\right) }
  Let \(c\), \(c'\), \(d\), and \(d'\) be four \(\C\)-classical constraints
 satisfying \(c \vdashc c'\) and \(d \vdashc
  d'\). For any constraint \(e\), all variables \(\bfvec x\) not free in \(d\),
  and all processes \(P\) and \(Q\), the following relations hold:
 \begin{center}\small%
\(\begin{array}{lll}
(1) ~~ \LOCALruleB && (2) ~~ \LOCALruleA \\
(3) ~~ \LOCALruleA && (4) ~~ \LOCALruleE \\
(5) ~~ \LOCALruleC  && (6) ~~ \LOCALruleH  \\
 (7) ~~ \LOCALruleD && (8) ~~ \LOCALruleI  
\end{array}
\)
\end{center}
\end{proposition}

The proof of the propositions relies on the following
lemma, that states a process emits classical constraints without
weaken itself.

\begin{lemma}
\label{lemma:classical}
Let \(\D\) and \(\E\) be two sets of constraints, \(P\) and \(P'\) two
processes and \(c\) a \(\C\)-classical constraint. If \(P \lraomc{c} P'\)
then \(P \equiv_\C P'\).
\end{lemma}

The may-testing relation
\(\simeq_{\C_c}\) coincides with an
equivalence used by Saras\-wat to connect operational and denotational
semantics of CC~\cite{SRP91popl}. Weaker versions of laws \((1)\) to
$(6)$ are proved indirectly for this relation.  Saraswat has also
defined a bisimulation semantics for CC~\cite{SR90popl}. The
bisimulation he proposed is strong (\ie\ it is based on {\small
  $\lramc{\alpha}$} instead of {\small $\Lramc{\alpha}$}), and is
therefore maybe too discriminative for an asynchronous framework such
as CC. For instance, none of the above laws,
except \((2)\), can hold for any reasonable notion of strong
bisimulation. This difference aside, Saraswat's bisimulation seems
still too discriminative. Indeed, on contrary to \(\cong_{\C_c\C_c}\),
it distinguishes processes like
\((\ask{x\!<\!1\!}{\!\tell{c}})\, \para\, (\ask{x \! < \!
  2\!}{\!\tell{c}})\) and \((\ask{x\!< \!2\!}{\!\tell{c}})\, \para\,
(\ask{x \!<\!  2\!}{\!\tell{c}})\) (where \(x< y\) is the usual
arithmetic inequality constraint), whereas there is no reasonable
justification to do so (in both strong and weak case).

\subsection{Observational equivalence relations for  CHR}

The Constraint Handling Rules (CHR) programming
language~\cite{Fruehwirth09cambridge} is a multiset rewriting language
over first-order terms with constraints over arbitrary mathematical
structures.
Initially introduced for programming constraint solvers,
CHR has evolved since to a programming language in its own right.

\subsubsection{Constraint Handling Rules Syntax}

The formalization of CHR assumes a language of \emph{built-in
  constraints} containing the equality \(=\), \(\false\), and \(\true\) over
some theory \(\CT\) and defines \emph{user-defined constraints} using a
different set of predicate symbols.
We require the non-logical axioms of \(\CT\) to be formulas of the
form \(\forall (\CC \rightarrow \exists Z.\DD)\), where both \(\CC\)
and \(\DD\) stand for possibly empty conjunctions of built-in
constraints.  Constraint theories satisfying such requirements
correspond to
Saraswat's {\em simple constraints systems} \citeNN{SRP91popl}.

A {\em CHR program} is a finite set of eponymous rules of
the form 
\((r \at \KK \backslash \HH \simplif \GG \mid \CC, \BB)\),
where \(\KK\), \(\HH\) are multisets of user-defined constraints,
called {\em kept head} and {\em removed head} respectively, \(\GG\) is
a conjunction of built-in constraints called {\em guard}, \(\CC\) is a
conjunction of built-in constraints, \(\BB\) is a multiset of
user-defined constraints, and \(r\) is an arbitrary identifier assumed
unique in the program called {\em rule name}.
Rules where both heads are empty are prohibited. The empty guard
\(\true\) can be omitted together with the symbol \(\mid\). Similarly,
empty keptheads can be omitted together with the
symbol \(\backslash\).
Propagation rules (\ie rules with empty removed head) can be written
using the alternative syntax: \(r \at \KK \propag \GG \mid \CC, \BB\).
A {\em state} is a tuple \(\state{\CC}{\EE}{X}\), where
\(\CC\) is a multiset of CHR constraints
, $\EE$
is a conjunction of built-in constraints,
and $X$ is a set of variables.

\subsubsection{From Constraints Handling Rules to Linear Logic CC}

In a recent paper, \citeN{Martinez10ppdp} has proposed a translation
from CHR to a subset of LCC (and vice versa), that preserves language
semantics with strong bisimilarity. This result allows us to
transpose straightforwardly our different notions of observational
equivalence to CHR.  To the best of our knowledge, it is the first
attempt to provide CHR with such equivalence techniques.

In \Cref{fig:translation:chr:lcc}, we recall Martinez's LCC interpretation
of basic CHR constructs.
\begin{table}
  \centering
\[\begin{array}{lrrl}
\text{Built-in const.} &   (c_1 \wedge \dots \wedge c_n)^\times &\!\!=\!\!& \oc  c_1  \otimes \dots \otimes !c_n\\
\text{CHR const.} &    (d_1 , \dots , d_n)^\times &\!\!=\!\!& d_1  \otimes \dots \otimes d_n\\
\text{Rules} &  r\at(\KK \backslash \HH \simplif \GG \mid \BB)^\times &\!\!=\!\!&
  \oc\ask[]{\KK ^\times\! \otimes \HH^\times\! \otimes
    \GG^\times}{\exists Y{\prth{\KK ^\times\!   \otimes \GG^\times\!
          \otimes \BB^\times}}}\\
\text{Program} & \{ r_1, \dots, r_n \}^\times &\!\!=\!\!& r_1 ^\times  \para \cdots \para r_n^\times  \\
\text{State} &  \state{\EE}{\CC}{X} ^ \circ &\!\!=\!\!& \exists ~(\EE ^\times\! \otimes \CC ^\times)
  \end{array}\]
\qquad where \( Y = \prth{\fv(\GG, \BB) \setminus \fv(\HH, \KK)}\) and \( Z = \prth{\fv(\EE, \CC) \setminus X}\). \hfill\null
  \caption{Translation from CHR to LCC}
  \label{fig:translation:chr:lcc}
\end{table}
A CHR state \(\sigma\) together with a CHR program \(\Prog\) are interpreted
as the process \((\sigma^\times \para \Prog^\times)\). The constraint
theory, \(\CT\), is translated using a standard translation of
intuitionistic logic into linear logic
. More precisely,
in the remainder of this section, \((\C,
\Vdashc)\) is the constraint system, where \(\C\) is built from the
built-in and CHR constraints and \(\Vdashc\) is defined by :
\((\forall(\CC \rightarrow \exists \DD)) \in \CT\) if and only
\(\CC^\times \Vdashccc \exists X \DD^\times\).

  Due to space limitation, we
 do not recall the
operational semantics of CHR, but use translations of CHR as
particular instances of LCC processes.  Thanks to Martinez's semantics
preservation theorem \citeNN{Martinez10ppdp}, we can do so without
loss of generality as long as the CHR
abstract semantics is concerned. In fact, we know that for any CHR
state \(\sigma\) and any CHR program \(\Prog\), \((\sigma^\times, \Prog^\times)
\Lramc{\tau} Q\) if and only if \(\sigma\) can be
rewritten by \(\Prog\) (\wrt\ CHR abstract semantics) to a state \(\sigma'\)
\st\ \(Q \equiv (\sigma'^\times, \Prog^\times)\). For the sake of
conciseness, we will write \(\sigma \mapsto_\Prog \sigma'\) for
\((\sigma^\times \para \Prog^\times ) \Lramc{\tau} (\sigma^\times \para
\Prog^\times)\).

\subsubsection{Confluence up to}

Confluence is an important property for CHR programs, which
  ensures that any computation for a goal results in the same final
  state (\ie\ modulo the structural equivalence \(\equivc\)) no matter
  which of the applicable rules are used. Here we propose a
  straightforward extension, called confluence up to, where structural
  equivalence is replaced by an observational one.  The
  resulting notion differs form the so-called observable confluence
  \cite{DSS07iclp} in the following sense: Observable confluence
  consists of proving that a program is confluent on an interesting
  subset of the states, while confluence up to consists of proving
  that a (possibly nowhere confluent) program is apparently confluent
  to an external observer.

\begin{definition}[Confluence up to]
  Let \(\D\) and \(\E\) be two sets of linear constraints.  A CHR
  program \(\Prog\) is {\em confluent up to \(\cong_{\D\E}\)} if whenever
  \(\sigma \mapsto_\Prog^* \sigma_1\) and \(\sigma \mapsto_\Prog^*
  \sigma_2\), there exist \(\sigma_1\) and \(\sigma_2\) such that
  \(\sigma_1 \mapsto_\Prog^* \sigma_1'\), \(\sigma_2\mapsto_\Prog^*
  \sigma_2'\), and \((\sigma_1'^\times\para \Prog^\times) \cong_{\D\E}
  (\sigma_2'^\times\para \Prog^\times)\). %
\end{definition}

The following proposition states that CHR transitions \wrt\ a
confluent program are not observable by any barbed congruences
obsevring only classical constraints.  The choice of limiting
observation to classical constraints makes sens since CHR programs are
usually embedded in a (host language) module that prohibs an external
observer synchronizing on internal CHR constraints; the observer can only
post CHR constraints using the module interface. As it is the case for
\cref{proposition:chr}, the proof relies on \cref{lemma:classical}.

\begin{proposition}
\label{proposition:chr:confluence}
Let \(\Cc\) be a set of \(\C\)-classical constraints and \(\D\) a
  set of linear constraints.  If \(\Prog\) is confluent up to
\(\cong_{\Cc\D}\) then \((\sigma^\times \para
\Prog^\times) \Lramc{\tau}\! P\) implies \((\sigma^\times\para \Prog^\times)\!
\cong_{\Cc\D}\! P\).
\end{proposition}

As corollary, we obtain that barbed congruences and may-testing
equivalences conincide when they observe only classical (\ie\
built-in) constraints. This supports the intuitive idea that a
  confluent program has the same meaning in the ``backtracking'' and
  the ``commited choice'' exuction paradigms -- bearing in mind that
  both relations are the respective instances of observation
  equivalences for these paradigms.

\begin{corollary}\label{corollary:chr:may}
  Let \(\Cc\) be a set of \(\C\)-classical constraints.  Let \(\Prog\) and \(\Prog'\) 
  be two CHR programs confluent up to \(\cong_{\Cc\D}\).  For all states
  \(\sigma\) and \(\sigma'\), \((\sigma^\times\para \Prog^\times)
  \simeq_{\Cc\D} (\sigma'^\times\para \Prog'^\times)\) if and only if
  \((\sigma^\times\para \Prog^\times) \cong_{\Cc\D} (\sigma'^\times\para
  \Prog'^\times)\)
\end{corollary}

\subsubsection{Application}

Observational equivalences are commonly used to prove correctness of a
realistic (or efficient) implementation \wrt\ a given
specification. See, for instance, numerous examples in Milner's book
\citeNN{Milner89prentice}. Here, we illustrate such a use in the
context of CHR.
For instance, let us assume given the following specification program
$\Prog_s$:
\begin{center}\small
$\begin{array}{ll} 
 \text{symmetry}& \;@\; \eq(x, y) \Longrightarrow \eq(y, x) \\
 \text{transitivity}& \;@\;  \eq(x, y), \eq(y, z)  \Longrightarrow \eq(y, z) \\
 \text{decompose}& \;@\; \eq(\fun(x_f,x_l,x_r), \fun(y_f, y_l, y_r))  \Longrightarrow x_f = y_f,  \eq(x_l, y_l) , \eq(x_r, y_r)
\end{array}$
\end{center}

One can be easily convinced that this program specifies a
Rational Terms (RT) solver limited to labelled binary trees: A binary
node is represented by a term {\small $\fun(x_f, x_l, x_r)$}, where
{\small $x_f$} is a label (or functor), and {\small $x_l$}, {\small
  $x_r$} are the left and right subtrees, respectively. %
Here, we aim at providing a program observationally equivalent to
$\Prog_s$ that is usable in practice.  As argued previously, since a CHR
solver is typically isolated in a host module, it is reasonable to
restrict the power of the observer such that it cannot observe CHR
constraints and can post only public (or exported) CHR
constraints. Hence, we choose {\small $\C_c$} and {\small $\C_c^\eq =
  \C_c \cup \C^\eq$} (where {\small $\C^\eq$} is the set of
constraints of the form {\small $\eq(s,t)$}) as input and output
filters, respectively.  Since CHR is a committed choice language, we
have to provide a program {\small $\Cc\C_c^\eq$}-barbed congruent
with {\small $\Prog_s$}.

A possible implementation for the RT problem has been proposed by
\citeN{Fruehwirth09cambridge}. This program uses extra-logical
constraints such as $var/1$.  Here we prefer writing pure programs,
since the status of the extra-logical constraints is not firmly
defined in the theoretical semantics.
For this reason, we propose the program {\small $\Prog_i$} given below.
To solve the problem, this program roughly emulates Prolog's unification
algorithm~\cite{Ait-kacy91wam} -- a constraint {\small $\eq(t,s)$}
encodes an equations to be solved, and a constraint {\small $x \ratr
  t$} encodes the unification (or the binding) of a variable {\small
  $x$} with a term {\small $t$}. We argue that $\Prog_i$ is more realistic than
{\small $\Prog_s$} since it terminates under the refined semantics of CHR
\cite{DSGH04iclp} -- which selects rules in the syntactical order
whereas $\Prog_s$ has no terminating derivation.
\begin{center}
\small
$
\begin{array}{ll}
  \text{reflex}& \;@\;\eq(x, x) \simplif \true. \\
  \text{decompose}& \;@\; \eq(\fun(x_f,x_l,x_r), \fun(y_f, y_l, y_r))  \simplif x_f = y_f,  \eq(x_l, y_l) , \eq(x_r, y_r).\\
  \text{orient} & \;@\; \eq(\fun(x_f,x_l,x_r), y) \simplif \eq(y, \fun(x_f,x_l,x_r)).\\
  \text{deref\_left}& \;@\;x \ratr z \backslash \eq(x, y) \simplif \eq(z, y) \\
 \text{deref\_right}& \;@\;y \ratr z \backslash \eq(x, y) \simplif \eq(x, z) \\ 
 \text{unif} & \;@\; \eq(x,y) \simplif x \ratr y.\\
\end{array}
$
\end{center}

Unfortunately, {\small $\Prog_i$} is not {\small $\Cc\C_c^\eq$}-barbed
congruent with the specification {\small $\Prog_s$}.  For instance, for
any {\small $\sigma_s$} \st\ {\small $\langle \eq(x, \fun(a, y, z)),
  \eq(x, \fun(a, y, z)), \true , \emptyset \rangle \mapsto_{\Prog_s}^*
  \sigma_s$}, we have {\small $\false \in \oac{\Cc}{\sigma_s^\times \para
  \Prog_s^\times}$}, but for $\sigma_i = \langle x \ratr \fun(a, y, z)),
x \ratr \fun(a, y, z)), \true, \emptyset \rangle$, we have {\small $
  \langle \eq(x, \fun(a, y, z)), \eq(x, \fun(a, y, z)), \true ,
  \emptyset \rangle \mapsto_{\Prog_i}^* \sigma_i$} and {\small $\false
  \notin \oac{\Cc}{\sigma_i^\times \para \Prog_i^\times}$.}  One simple idea
to circumvent this problem is to ``complete'' $P_i$ \cite{AF98cp}
(\ie\ to make it confluent by adding new rules). For instance, one can
add at the end of $\Prog_i$ the following rules.  Intuitively these
rules ``repair'' states that do not respect the binding invariant
(\ie\ only variables are bound, only once, and not to themselves),
which is normally preserved by the refined semantics -- as far as the
observer do not performed built-in unification.
\begin{center}\small$
  \begin{array}[c]{ll}
   \text{repair}_1 & \;@\; \fun(x_f, x_l, x_r) \ratr y \simplif \eq(y,  \fun(x_f, x_l, x_r)).\\
   \text{repair}_2 & \;@\; x \ratr y \backslash x \ratr z \simplif \eq(x, z).\\
   \text{repair}_3 & \;@\; x \ratr x \simplif \true.
  \end{array}
$
\end{center}

The resulting program {\small $\Prog_i^\star$} is confluent up to {\small
  $\cong_{\Cc\C_c^\eq}$} and {\small $\Cc\C_c^\eq$}-barbed congruent
with {\small $\Prog_s$}. The proof can be sketched as follows:
Assume the function {\small $(\, )^\eq$} defined on atomic constraints
as {\small $c^\eq = eq(t,s)$} if {\small $c$} is of the form {\small
  $(t \ratr s)$}, or {\small $c^\eq = c$} otherwise. Consider the
relation {\small $\R = \{ (P^\times \para c),(P^\times \para c^\eq) |
  c \in \C\}$} where {\small $( )^\eq$} is extended to non-atomic
constraints in the straightforward way. First, we prove by coindutive
reasoning on the transition from {\small $(P^\times \para c)$} that
{\small $\R$} is a {\small $\Cc\C_c^\eq$}-bismulation, or thanks to
\cref{theorem:bisimulation} a {\small $\Cc\C_c^\eq$}-barbed
congruence. Then, by using a straightforward extension of strong
confluence for abstract rewritting system~\cite{Huet80jacm}, we show
that {\small $\Prog_i^\star$} is confluent up to {\small $\R$}, \ie,
confluent up to $\cong_{\Cc\C_c^\eq}$.  Finally, we demonstrate by a
structural induction on the {\small $\Cc\C_c^\eq$}-contexts that
{\small $\Prog_i^\star \simeq_{\Cc\C_c^\eq} \Prog_s$}, or thanks to
\cref{corollary:chr:may}, {\small $\Prog_i^\star \cong_{\Cc\C_c^\eq}
  \Prog_s$}.

Therefore, {\small $\Prog_i^\star$} is a correct implementation of
{\small $\Prog_s$}.  But, since we have proven that {\small $\Prog_i^\star$}
is also confluent, we know it can be interpreted under any rule
selection strategy (in particular, under the one of the refined
semantics) without loosing completeness. For this reason, and because
the ``repair'' rules are never called under the refined semantics as
long as the observer does not performed built-in unification, {\small
  $\Prog_i$} interpreted in the refined semantics is also a correct
implementation of {\small $\Prog_s$}. Note that Fr\"uhwirth's RT also cannot
deal with built-in unifications because of the non-monotonicity of
extra-logical constraints, while $\Prog_i^\star$ can.

To the best of our knowledge, the only existing notion of equivalence
for CHR programs that can be related to observation equivalences is
the so-called operational equivalence~\cite{AF99cp}. This notion means
that given two confluent and terminating programs, the computation of
a query in both programs terminates in the same state. Nonetheless, we
argue that observable equivalences are more general than operational
equivalence, since they can also be applied to programs such as
$\Prog_i^*$ which is non-terminating, non-confluent, and whose final
states contain distinct CHR constraints

\section{Conclusion}

In the first part of this paper we have defined and investigated a
structural operational semantics for LCC with quantified ask and
replication.  In light of this new semantics, we have proposed and
studied several observational equivalence
relations.  To the best of our knowledge, it is the first attempt to
provide LCC with such tools, even though it was identified early on as
a worthwhile goal of investigation by Ruet~\cite{RF97csl}.

In the second part of this paper, we related LCC and its observational
equivalence to asynchronous process and CC frameworks. In particular,
we have shown that the asynchronous $\pi$-calculi can be viewed as
subcalculi of LCC. We have shown, moreover, that some of the usual
observational equivalence relations defined for this calculus are
particular instances of the ones we have defined for LCC. 
Finally, we have shown that LCC observational equivalences can be
transposed straightforwardly to classical CC and CHR.  We have
demonstrated some interesting properties of the resulting
equivalences. In particular, we have studied the
relation between barbed-congruence and confluence of
CHR programs. We illustrated also how
barbed-congruence can be used to prove realistic
implementation constraint solvers \wrt\ a simple specification.

An immediate further work could be to  investigate the
properties of the observational equivalence relations presented
here. For instance, establishing sufficient
conditions 
to ensure that observational equivalences are full congruences would
be interesting.  It should also be worthwhile to
formally compare LCC with more exotic asynchronous calculi, such as
hybrid process calculi with constraints
\cite{DRV98clei,PV98lics,GP00cl,BM07esop} or extended calculi with
security primitives \cite{AFG00popl}, where the linear constraint
system would play a more prominent role.  Finally, the further
investigation of CHR bisimulation seems promising.

 \section*{Acknowledgements}

  The research leading to these results has received funding from the
  Madrid Regional Government under the CM project P2009/TIC/1465
  (PROMETIDOS), the Spanish Ministry of Science under the MEC project
  TIN-2008-05624 {\em DOVES}, and the EU Seventh Framework
  Programme FP/2007-2013 under grant agreement 215483 (S-CUBE).

We thank the reviewers for their helpful and
constructive comments.

\bibliographystyle{acmtrans}
\bibliography{contraintes}

\begin{thebibliography}{}

\bibitem[\protect\citeauthoryear{Abadi, Fournet, and Gonthier}{Abadi
  et~al\mbox{.}}{2000}]{AFG00popl}
{\sc Abadi, M.}, {\sc Fournet, C.}, {\sc and} {\sc Gonthier, G.} 2000.
\newblock Authentication primitives and their compilation.
\newblock In {\em POPL}. ACM PRESS, 302--315.

\bibitem[\protect\citeauthoryear{Abdennadher and Fr\"{u}hwirth}{Abdennadher and
  Fr\"{u}hwirth}{1998}]{AF98cp}
{\sc Abdennadher, S.} {\sc and} {\sc Fr\"{u}hwirth, T.} 1998.
\newblock On completion of constraint handling rules.
\newblock In {\em CP}. LNCS, vol. 1520. Springer, 25--39.

\bibitem[\protect\citeauthoryear{Abdennadher and Fr{\"u}hwirth}{Abdennadher and
  Fr{\"u}hwirth}{1999}]{AF99cp}
{\sc Abdennadher, S.} {\sc and} {\sc Fr{\"u}hwirth, T.~W.} 1999.
\newblock Operational equivalence of chr programs and constraints.
\newblock In {\em CP}. LNCS, vol. 1713. Springer, 43--57.

\bibitem[\protect\citeauthoryear{A{\"\i}t-Kaci}{A{\"\i}t-Kaci}{1991}]{Ait-kacy91wam}
{\sc A{\"\i}t-Kaci, H.} 1991.
\newblock {\em Warren's Abstract Machine, A Tutorial Reconstruction}.
\newblock Logic Programming. MIT Press.

\bibitem[\protect\citeauthoryear{Amadio, Castellani, and Sangiorgi}{Amadio
  et~al\mbox{.}}{1998}]{ACS98tcs}
{\sc Amadio, R.~M.}, {\sc Castellani, I.}, {\sc and} {\sc Sangiorgi, D.} 1998.
\newblock On bisimulations for the asynchronous pi-calculus.
\newblock {\em Theor. Comput. Sci.\/}~{\em 195,\/}~2, 291--324.

\bibitem[\protect\citeauthoryear{Berry and Boudol}{Berry and
  Boudol}{1992}]{BB92tcs}
{\sc Berry, G.} {\sc and} {\sc Boudol, G.} 1992.
\newblock The chemical abstract machine.
\newblock {\em Theor. Comput. Sci.\/}~{\em 96,\/}~1, 217--248.

\bibitem[\protect\citeauthoryear{Best, de~Boer, and Palamidessi}{Best
  et~al\mbox{.}}{1997}]{BBP97coord}
{\sc Best, E.}, {\sc de~Boer, F.}, {\sc and} {\sc Palamidessi, C.} 1997.
\newblock Partial order and {SOS} semantics for linear constraint programs.
\newblock In {\em COORDINATION}. LNCS, vol. 1282. Springer, 256--273.

\bibitem[\protect\citeauthoryear{Buscemi and Montanari}{Buscemi and
  Montanari}{2007}]{BM07esop}
{\sc Buscemi, M.~G.} {\sc and} {\sc Montanari, U.} 2007.
\newblock {CC-P}i: A constraint-based language for specifying service level
  agreements.
\newblock In {\em ESOP}. LNCS, vol. 4421. Springer, 18--32.

\bibitem[\protect\citeauthoryear{D\'{\i}az, Rueda, and Valencia}{D\'{\i}az
  et~al\mbox{.}}{1998}]{DRV98clei}
{\sc D\'{\i}az, J.~F.}, {\sc Rueda, C.}, {\sc and} {\sc Valencia, F.~D.} 1998.
\newblock Pi+- calculus: A calculus for concurrent processes with constraints.
\newblock {\em CLEI Electron. J.\/}~{\em 1,\/}~2.

\bibitem[\protect\citeauthoryear{Duck, Stuckey, {Garc{\'i}a de la Banda}, and
  Christian}{Duck et~al\mbox{.}}{2004}]{DSGH04iclp}
{\sc Duck, G.~J.}, {\sc Stuckey, P.~J.}, {\sc {Garc{\'i}a de la Banda}, M.},
  {\sc and} {\sc Christian, H.} 2004.
\newblock The refined operational semantics of {C}onstraint {H}andling {R}ules.
\newblock In {\em ICLP}. LNCS, vol. 3132. Springer, 90--104.

\bibitem[\protect\citeauthoryear{Duck, Stuckey, and Sulzmann}{Duck
  et~al\mbox{.}}{2007}]{DSS07iclp}
{\sc Duck, G.~J.}, {\sc Stuckey, P.~J.}, {\sc and} {\sc Sulzmann, M.} 2007.
\newblock Observable confluence for constraint handling rules.
\newblock In {\em ICLP}. LNCS, vol. 4670. Springer, 224--239.

\bibitem[\protect\citeauthoryear{Fages, Ruet, and Soliman}{Fages
  et~al\mbox{.}}{2001}]{FRS01ic}
{\sc Fages, F.}, {\sc Ruet, P.}, {\sc and} {\sc Soliman, S.} 2001.
\newblock Linear concurrent constraint programming: Operational and phase
  semantics.
\newblock {\em Inf. Comput.\/}~{\em 165,\/}~1, 14--41.

\bibitem[\protect\citeauthoryear{Fournet and Gonthier}{Fournet and
  Gonthier}{2005}]{FG05jlap}
{\sc Fournet, C.} {\sc and} {\sc Gonthier, G.} 2005.
\newblock A hierarchy of equivalences for asynchronous calculi.
\newblock {\em J. Log. Algebr. Program.\/}~{\em 63,\/}~1, 131--173.

\bibitem[\protect\citeauthoryear{Fr\"{u}hwirth}{Fr\"{u}hwirth}{2009}]{Fruehwirth09cambridge}
{\sc Fr\"{u}hwirth, T.} 2009.
\newblock {\em Constraint Handling Rules}.
\newblock Cambridge University Press.

\bibitem[\protect\citeauthoryear{Gilbert and Palamidessi}{Gilbert and
  Palamidessi}{2000}]{GP00cl}
{\sc Gilbert, D.} {\sc and} {\sc Palamidessi, C.} 2000.
\newblock Concurrent constraint programming with process mobility.
\newblock In {\em Computational Logic}. LNCS, vol. 1861. Springer, 463--477.

\bibitem[\protect\citeauthoryear{Girard}{Girard}{1987}]{Girard87tcs}
{\sc Girard, J.-Y.} 1987.
\newblock Linear logic.
\newblock {\em Theoretical Computer Science\/}~{\em 50(1)}.

\bibitem[\protect\citeauthoryear{Haemmerl\'{e}}{Haemmerl\'{e}}{2011}]{Haemmerle11clipB}
{\sc Haemmerl\'{e}, R.} 2011.
\newblock Toward observational equivalences for linear logic concurrent
  constraint languages.
\newblock Tech. Rep. CLIP5/2011, Technical University of Madrid (UPM).

\bibitem[\protect\citeauthoryear{Haemmerl{\'e}, Fages, and
  Soliman}{Haemmerl{\'e} et~al\mbox{.}}{2007}]{HFS07fsttcs}
{\sc Haemmerl{\'e}, R.}, {\sc Fages, F.}, {\sc and} {\sc Soliman, S.} 2007.
\newblock Closures and modules within linear logic concurrent constraint
  programming.
\newblock In {\em FSTTCS}. LNCS, vol. 4855. Springer, 554--556.

\bibitem[\protect\citeauthoryear{Honda and Yoshida}{Honda and
  Yoshida}{1995}]{HY95tcs}
{\sc Honda, K.} {\sc and} {\sc Yoshida, N.} 1995.
\newblock On reduction-based process semantics.
\newblock {\em Theor. Comput. Sci.\/}~{\em 151,\/}~2, 437--486.

\bibitem[\protect\citeauthoryear{Huet}{Huet}{1980}]{Huet80jacm}
{\sc Huet, G.} 1980.
\newblock Confluent reductions: Abstract properties and applications to term
  rewriting systems: Abstract properties and applications to term rewriting
  systems.
\newblock {\em Journal of the ACM\/}~{\em 27,\/}~4 (Oct.), 797--821.

\bibitem[\protect\citeauthoryear{Jaffar and Lassez}{Jaffar and
  Lassez}{1987}]{JL87popl}
{\sc Jaffar, J.} {\sc and} {\sc Lassez, J.-L.} 1987.
\newblock Constraint logic programming.
\newblock In {\em Proceedings of the 14th ACM Symposium on Principles of
  Programming Languages, Munich, Germany}. ACM, 111--119.

\bibitem[\protect\citeauthoryear{Kahn and Saraswat}{Kahn and
  Saraswat}{1990}]{KS90oopsla}
{\sc Kahn, K.~M.} {\sc and} {\sc Saraswat, V.~A.} 1990.
\newblock Actors as a special case of concurrent constraint programming.
\newblock In {\em OOPSLA/ECOOP}. 57--66.

\bibitem[\protect\citeauthoryear{Laneve and Montanari}{Laneve and
  Montanari}{1992}]{LM92mfcs}
{\sc Laneve, C.} {\sc and} {\sc Montanari, U.} 1992.
\newblock Mobility in the {CC}-paradigm.
\newblock In {\em MFCS}. LNCS, vol. 629. Springer, 336--345.

\bibitem[\protect\citeauthoryear{Maher}{Maher}{1987}]{Maher87iclp}
{\sc Maher, M.~J.} 1987.
\newblock Logic semantics for a class of committed-choice programs.
\newblock In {\em Proceedings of ICLP'87, International Conference on Logic
  Programming}. MIT Press.

\bibitem[\protect\citeauthoryear{Martinez}{Martinez}{2010}]{Martinez10ppdp}
{\sc Martinez, T.} 2010.
\newblock Semantics-preserving translations between linear concurrent
  constraint programming and constraint handling rules.
\newblock In {\em PPDP}. ACM PRESS, 57--66.

\bibitem[\protect\citeauthoryear{Milner}{Milner}{1989}]{Milner89prentice}
{\sc Milner, R.} 1989.
\newblock {\em Communication and Concurrency}.
\newblock Prentice Hall.

\bibitem[\protect\citeauthoryear{Milner and Sangiorgi}{Milner and
  Sangiorgi}{1992}]{MS92icalp}
{\sc Milner, R.} {\sc and} {\sc Sangiorgi, D.} 1992.
\newblock Barbed bisimulation.
\newblock In {\em ICALP}. LNCS, vol. 623. Springer, 685--695.

\bibitem[\protect\citeauthoryear{Nicola and Hennessy}{Nicola and
  Hennessy}{1984}]{DNH84tcs}
{\sc Nicola, R.~D.} {\sc and} {\sc Hennessy, M.} 1984.
\newblock Testing equivalences for processes.
\newblock {\em Theor. Comput. Sci.\/}~{\em 34}, 83--133.

\bibitem[\protect\citeauthoryear{Parrow and Victor}{Parrow and
  Victor}{1998}]{PV98lics}
{\sc Parrow, J.} {\sc and} {\sc Victor, B.} 1998.
\newblock The fusion calculus: Expressiveness and symmetry in mobile processes.
\newblock In {\em LICS}. {IEEE}, 176--185.

\bibitem[\protect\citeauthoryear{Ruet and Fages}{Ruet and
  Fages}{1997}]{RF97csl}
{\sc Ruet, P.} {\sc and} {\sc Fages, F.} 1997.
\newblock Concurrent constraint programming and mixed non-commutative linear
  logic.
\newblock In {\em CSL}. LNCS, vol. 1414. Springer, 406--423.

\bibitem[\protect\citeauthoryear{Saraswat and Lincoln}{Saraswat and
  Lincoln}{1992}]{SL92xerox}
{\sc Saraswat, V.~A.} {\sc and} {\sc Lincoln, P.} 1992.
\newblock Higher-order linear concurrent constraint programming.
\newblock Tech. rep., Xerox Parc.

\bibitem[\protect\citeauthoryear{Saraswat and Rinard}{Saraswat and
  Rinard}{1990}]{SR90popl}
{\sc Saraswat, V.~A.} {\sc and} {\sc Rinard, M.~C.} 1990.
\newblock {C}oncurrent constraint programming.
\newblock In {\em POPL}. ACM PRESS, 232--245.

\bibitem[\protect\citeauthoryear{Saraswat, Rinard, and Panangaden}{Saraswat
  et~al\mbox{.}}{1991}]{SRP91popl}
{\sc Saraswat, V.~A.}, {\sc Rinard, M.~C.}, {\sc and} {\sc Panangaden, P.}
  1991.
\newblock Semantic foundations of concurrent constraint programming.
\newblock In {\em POPL}. ACM.

\bibitem[\protect\citeauthoryear{Soliman}{Soliman}{2003}]{soliman03inria}
{\sc Soliman, S.} 2003.
\newblock Pi-calculus and {LCC}, a space odyssey.
\newblock Research Report 4855, INRIA.

\bibitem[\protect\citeauthoryear{Tamura}{Tamura}{1998}]{Tamura98Kobe}
{\sc Tamura, N.} 1998.
\newblock {\em User's Guide of a Linear Logic Theorem Prover}.
\newblock Kobe University.
\newblock {\tt http://bach.istc.kobe-u.ac.jp/llprover/}.

\end{thebibliography}

\end{document}